\documentclass[
 reprint,
 amsmath,amssymb,
 aps,nobibnotes,
 prl,superscriptaddress
]{revtex4-1}
\usepackage[colorlinks=true,linkcolor=blue,citecolor=blue,urlcolor=blue]{hyperref}

\begin{document}

\title{Searches for new sources of CP violation using molecules as quantum sensors \vspace{1cm}}
\author{N. R. Hutzler$^*$}
\affiliation{California Institute of Technology, Pasadena, CA, USA}
\email{hutzler@caltech.edu}
\author{A. Borschevsky}
\affiliation{University of Groningen, Groningen, The Netherlands}
\author{D. Budker}
\affiliation{Helmholtz-Institute and Johannes Gutenberg-Universit\"{a}t, Mainz, Germany}
\affiliation{University of California, Berkeley, CA, USA}
\author{D. DeMille}
\affiliation{University of Chicago, Chicago, IL, USA}
\affiliation{Argonne National Laboratory, Argonne, IL, USA}
\author{V. V. Flambaum}
\affiliation{University of New South Wales, Sydney, New South Wales, Australia}
\author{G. Gabrielse}
\affiliation{Northwestern University, Evanston, IL, USA}
\author{R. F. Garcia Ruiz}
\affiliation{Massachusetts Institute of Technology, Cambridge, MA, USA}
\author{A. M. Jayich}
\affiliation{University of California, Santa Barbara, CA, USA}
\author{L. A. Orozco}
\affiliation{Joint Quantum Institute, National Institute of Standard and Technology and
University of Maryland, College Park, MD, USA}
\author{M. Ramsey-Musolf}
\affiliation{University of Massachusetts, Amherst, MA, USA}
\author{M. Reece}
\affiliation{Harvard University, Cambridge, MA, USA}
\author{M. S. Safronova}
\affiliation{University of Delaware, Newark, Delaware, USA}
\affiliation{Joint Quantum Institute, National Institute of Standard and Technology and
University of Maryland, College Park, MD, USA}
\author{J. T. Singh}
\affiliation{National Superconducting Cyclotron Laboratory and Michigan State University, East Lansing, Michigan, USA}
\author{M. R. Tarbutt}
\affiliation{Imperial College London, London, UK}
\author{T. Zelevinsky}
\affiliation{Columbia University, New York, NY, USA}

\date{\today}
\;\;

\vspace{3cm}

\maketitle

\newpage

\textbf{\emph{Introduction}}

There is compelling motivation, provided by the baryon asymmetry of the universe and the strong CP problem, to mount improved searches for new sources of CP violation (CPV). New quantum fields beyond those in the Standard Model frequently add new CPV phases. Moreover, prior searches for new CPV phenomena---experiments to detect anomalies in kaon mixing or particle electric dipole moments (EDMs)---are already sensitive to generic, new CPV physics at scales of $\gtrsim 10$ TeV, and even higher for certain models,  far beyond the direct reach of the LHC. Particle EDMs stand out as systems for future efforts in this direction: they are effectively background-free probes for new physics, since the Standard Model predicted values are many orders of magnitude below current limits \cite{Safronova2018,Chupp2019}.

There has been recent and very rapid progress in sensitivity to the electron EDM. The improved limits leveraged strong amplification of observable energy shifts due to the eEDM when electrons are bound in polar molecules. Though molecular amplification of CPV has been understood for decades, only recently were experimental techniques developed that create and control molecules at the level of sophistication needed for good statistical sensitivity and suppression of systematic errors.

We argue here that \textbf{new molecule-based searches for new CPV physics are poised to make many orders of magnitude of improvements in the coming decade and beyond}. These experiments will probe not only the electron EDM, but also hadronic CPV phenomena such as nuclear Schiff moments and magnetic quadrupole moments \cite{Chupp2019}. Generic sensitivity to flavor-neutral sources of both leptonic and hadronic CPV will be sufficient to probe scales $\gtrsim 100$ TeV, and flavor-changing CPV at scales of $\gtrsim 1000$ TeV---surpassing even the long-standing bounds from kaon decays. This anticipated progress will leverage ongoing, extremely rapid development of methods for creating, cooling, trapping, and controlling molecules---methods driven in large part by possible applications of ultracold molecules in Quantum Information Science. It will also make use of conceptual advances to identify optimal molecular species for different types of new CPV physics.

\textbf{\emph{Why molecules?}}

Atoms and neutrons have been sensitive platforms for precision measurements of low-energy CPV for many decades \cite{Ramsey1957,Safronova2018,Chupp2019}.  Molecules, on the other hand, have only recently surpassed the sensitivity of atomic measurements in a few key areas. Compared to atoms, the larger polarizability and internal fields of molecules results in up to a thousand-fold increase in sensitivity to fundamental symmetry violations such as electron EDMs, CPV electron-nucleon interactions, and hadronic CPV \cite{Safronova2018,DeMille2017,Chupp2019,Cairncross2019,Hutzler2020}. Though the complexity of molecules makes them challenging to control, the inherent advantages of their complex structure far outweighs this drawback.  Indeed, molecules are already the most sensitive probe of the eEDM \cite{Hudson2011,Baron2014,Cairncross2017,ACME2018}, having recently overtaken atomic experiments \cite{Regan2002} and \textbf{improved their sensitivity by two orders of magnitude in just over one decade}. Hadronic CPV is still mostly constrained by $^{199}$Hg \cite{Graner2016} and neutrons \cite{Abel2020}, though molecules have similar advantages in this area and many experiments have recently commenced.

Additionally, the molecular experiments are new and there are still orders of magnitude yet to be gained in sensitivity. A rough figure of merit for these searches, which rely on coherent precession of electrons or nuclei in internal molecular fields (analogous to the precession of free neutrons in external electromagnetic fields for nEDM searches) is (coherence time)$\times$(amplification)$\times$(count rate)$^{1/2}$. We discuss how each of these areas has significant untapped experimental potential in molecular systems.

\textbf{\emph{Pathways to improved experiments}}

Coherence times for beam experiments can be improved by using a longer beam line or a slower beam. Trapping molecules yields the longest possible coherence time, up to (and beyond) $10^3$ times longer than current beam experiments. Count rates can be increased by beam cooling and focusing, improvements in detection efficiency, improved trapping techniques, and brighter molecular sources. Intrinsic amplification can be achieved by using heavy species with large internal electromagnetic fields, and hadronic CPV sensitivity can be increased by using species with deformed nuclei. Every one of these areas is either being improved or implemented in experiments both ongoing and proposed.

Three molecules have been used to set an eEDM limit more sensitive than that of any atom, and are currently probing the few TeV scale for generic CPV physics in multiple sectors \cite{Engel2013,Cesarotti2019,Chupp2019}, and up to a thousand TeV for certain models \cite{Fuyuto2019}. The ACME ThO experiment \cite{ACME2018}, which currently has the most sensitive limit of $|d_e|< 1.1\times10^{-29}~e$~cm, is improving coherence time with a longer beam line, and increasing count rates through molecular flux and detection efficiency. The JILA HfF$^+$ ion trap experiment, which has already performed an EDM search with a coherence time far beyond that available to beams \cite{Cairncross2017}, has increased this coherence time even further \cite{Zhou2020} and is building a new apparatus to increase count rates, and has demonstrated the suitability of a species with an even larger sensitivity, ThF$^+$ \cite{Gresh2016}. The Imperial College YbF experiment \cite{Hudson2011}, which was the first to overcome the limit set by the atomic Tl experiment \cite{Regan2002}, has implemented a number of improvements to both molecular preparation and readout efficiency \cite{Ho2020}. Since each of these experiments was statistics-limited in their most recent result, there is considerable room for improvement. These improvements are in parallel to atomic eEDM experiments with Cs \cite{Zhu2013} and Fr \cite{Wundt2012,Inoue2014}. Other experiments are also underway, including those with BaF \cite{Aggarwal2018}, matrix-isolated molecules \cite{Vutha2018}, and advanced NMR techniques \cite{Budker2014,Eills2017,Wu2018,Blanchard2020}.

Experiments are also under construction with the goal of searching for hadronic CPV by leveraging the advantages which allowed molecular eEDM searches to become the most sensitive. These include the CENTReX nuclear Schiff moment search in a beam of TlF \cite{Hunter2012}, a nuclear magnetic quadrupole moment search in a beam of $^{173}$YbOH \cite{Kozyryev2017PolyEDM,Maison2019,Denis2020,Jadbabaie2020}, and experiments with radioactive RaF \cite{GarciaRuiz2020} and RaOCH$_3^+$\cite{Fan2020,Yu2020} discussed in a later section. These experiments are in parallel to improvements of existing atomic searches with $^{199}$Hg \cite{Graner2016}, $^{225}$Ra \cite{Parker2015,Bishof2016}, and $^{129}$Xe \cite{Sachdeva2019,Allmendinger2019}, and development of new experiments such as matrix-isolated $^{229}$Pa \cite{Singh2019}.  Note that both leptonic and hadronic CPV searches require multiple experiments with different systems to obtain robust bounds, as these effects can arise from multiple sources \cite{Chupp2015}.

\textbf{\emph{Advanced cooling methods}}

Laser cooling has been one of the main drivers of the tremendous quantum advances in the world of atomic physics, such as the atomic clocks now reaching unprecedented $<10^{-18}$ fractional uncertainty \cite{Oelker2019,Brewer2019}.  Implementing these advances in molecules sensitive to fundamental symmetry violations will result in orders of magnitude improvements.  Since the first laser cooling of a molecule in 2010 \cite{Shuman2010}, the field has advanced rapidly \cite{Tarbutt2018} and has resulted in several groups having directly cooled and trapped molecules at ultracold temperatures \cite{McCarron2018SrF,Caldwell2019,Anderegg2019,Ding2020}.  Several experiments are underway using laser-coolable molecules, including YbF \cite{Tarbutt2013,Lim2018}, BaF \cite{Aggarwal2018}, $^{174}$YbOH, \cite{Kozyryev2017PolyEDM,Denis2019,Prasannaa2019,Gaul2020,Augenbraun2020YbOH}, $^{173}$YbOH \cite{Kozyryev2017PolyEDM,Maison2019,Denis2020}, and TlF \cite{Cho1991,Hunter2012}. Shorter-term gains can come from beam slowing and cooling to increase count rates and coherence times. Longer-term and even more significant gains can come from trapping to achieve very long coherence times. Molecules can also be assembled from ultracold atoms \cite{Ni2008}, thereby creating them in a trap directly, and there are a number of candidate species \cite{Meyer2009,Sunaga2019} with sensitivity to CPV, such as AgRa \cite{AgRa,Sunaga2019}.

\textbf{\emph{Radioactive molecules}}

Heavy nuclei with static octopole deformations, such as Fr, Ra, Th, Pa, and others, can have hadronic CP-violation sensitivity enhancements up to a thousandfold larger than spherical nuclei \cite{Auerbach1996,Dobaczewski2005,Parker2015,Dobaczewski2018,Flambaum2019Schiff}. Combined with relativistic enhancements from their high mass, molecular species with deformed nuclei can be up to $10^6$ times more intrinsically sensitive \cite{Sushkov1985,Flambaum2019Schiff} than the current atomic Hg \cite{Graner2016}, which is the most sensitive atomic or molecular hadronic CPV experiment. Radium is of particular interest; it has a well-studied nuclear deformation \cite{Gaffney2013,Butler2020}, both the atom and many radium-containing molecules can be laser cooled \cite{Parker2015,Isaev2010,Kudashov2014,Isaev2017RaOH}, and atomic $^{225}$Ra is the subject of an EDM experiment at ANL \cite{Parker2015,Bishof2016}. RaF was recently spectroscopically studied \cite{GarciaRuiz2020}, and along with polyatomic analogues offer laser cooling and extreme sensitivity to hadronic symmetry violations. RaOCH$_3^+$, which was recently synthesized, trapped and cooled in an ion trap \cite{Fan2020}, offers similar sensitivity with the possibility for an experiment with advanced ion control techniques \cite{Fan2020,Yu2020,Cairncross2017,Chou2017}.

Molecules containing other nuclei are also of interest; many heavy nuclei such as Eu, Ac, Th, and others have longer lifetimes than the $^{225}$Ra isotope needed for a hadronic CPV search, yet have comparable sensitivity \cite{Flambaum2019Schiff,Skripnikov2020,Flambaum2020Schiff}. $^{229}$Pa is purported to possess and anomalously small splitting between opposite parity states \cite{Ahmad1982}, resulting in a factor of $\sim$40 further enhancement compared to radium \cite{Flambaum2008,Singh2019,Flambaum2020Schiff}, though with considerable nuclear structure uncertainties that must be addressed through further experiments.

\textbf{\emph{Advanced quantum control}}

The molecular CPV experiments discussed here rely on quantum superpositions and quantum control techniques for measurement.  These techniques are analogous to those used in quantum information science, and could therefore benefit from this rapidly-advancing field \cite{Cloet2019}. Far-future prospects include using entanglement-based squeezing to provide significant gains in sensitivity \cite{Hosten2016}, in addition to those discussed here.  However, this will require large, high-density samples at very low temperatures, and development of suitable measurement protocols.

\textbf{\emph{Outlook}}

Molecules are sensitive to a very wide range of fundamental physics, far beyond what is discussed here, and present opportunities to search for well-motivated new physics at scales accessible to few other kinds of experiments  \cite{Safronova2018,DeMille2017,Chupp2019,Cairncross2019,Hutzler2020}. They have already proven to be sensitive probes for CPV, and offer a realistic prospect for orders-of-magnitude increases in the coming decade and beyond. These rapid advances have been driven by new technologies to create, cool, and control complex species, and will continue to move forward in tandem with quantum information science.

While these experiments have the advantage of being relatively small (often $\lesssim$ 10 people) and inexpensive (often $\lesssim\$$10 M), they are increasing in scale and complexity as they advance to next-generation searches. Many of the new techniques proposed or under development require sustained R\&D budgets and theory support to enable exploration of multiple approaches, and support over multiple experimental generations to realize them. The field moves very rapidly, and requires a fair amount of risk tolerance, but has proven that it can deliver new results from a variety of new approaches.

\vspace{-1mm}

\bibliography{ref}

\end{document}